\documentclass{aastex}

\usepackage{psfig,emulateapj5}

\newcommand{\lsim}{\raisebox{0.3mm}{\em $\, <$} \hspace{-3.3mm}
\raisebox{-1.8mm}{\em $\sim \,$}}
\newcommand{\gsim}{\raisebox{0.3mm}{\em $\, >$} \hspace{-3.3mm}
\raisebox{-1.8mm}{\em $\sim \,$}}

\shorttitle{
SED in Supercritical Disk
through Photon-trapping Effects
}
\shortauthors{Ohsuga, Mineshige, Watarai}

\begin{document}


\title
{Spectral Energy Distribution in Supercritical Disk Accretion Flows 
through Photon-trapping Effects}


\author{K. Ohsuga, S. Mineshige, and K. Watarai}
\affil{Yukawa Institute for Theoretical Physics, Kyoto University,
Kyoto 606-8502, Japan}




\begin{abstract}
We investigate the spectral energy distribution (SED)
of the supercritical disk accretion flows around black holes
by solving the multi-frequency zeroth moment equation of the radiation transfer
equation 
under the flux-limited diffusion approximation 
as well as the energy equation of gas.
Special attention is paid to the photon-trapping effects,
the effects that photons are trapped within accretion flow
and are swallowed by a black hole, with little being radiated away.
It is found that 
when flow luminosity is below about twice the Eddington luminosity
($L\lsim 2L_{\rm E}$)
photon trapping is ineffective and
the SED shifts to the higher-energy side
as $L$ increases.
The peak frequency at which the SED reaches its peak becomes at maximum 
three times higher than
that given by the standard-disk model,
which might resolve the so-called 
\lq too hot accretion disk problem.'
When $L\gsim 2L_{\rm E}$,
in contrast the peak frequency of the SED tends to decrease
with increase of mass-accretion rate.
This is due to enhanced photon-trapping.
Since high-energy photons are generated near the equatorial
plane,
they can be more effectively trapped in flow
than low-energy ones and, hence, 
the high energy part of radiation is suppressed.
Finally, the observed time variation of IC 342 S1,
which is an ultraluminous X-ray source,
in the X-ray HR diagram
can be explained by the modulation of the mass accretion rate.
In comparison with the observational data,
we estimate the mass of black hole in IC 342 S1
is $\sim 100M_\odot$.
\end{abstract}



\section{INTRODUCTION}
The standard-disk model proposed by Shakura \& Sunyaev (1973)
has had some success in describing optically thick accretion flow structure
in luminous objects, such as X-ray binary sources in the soft state.
However, the standard-disk model is constructed under
number of approximations and its treatment is only justified 
when the mass-accretion rate, $\dot{M}$,
is less than the critical mass-accretion rate,
$\dot{M}_{\rm crit}$,
where the critical mass-accretion rate is given by
$\dot{M}_{\rm crit} \equiv L_{\rm E}/c^2$
with the Eddington luminosity,
$L_{\rm E}=4\pi cGMm_{\rm p}/\sigma_{\rm T}$,
where $c$, $M$, $m_{\rm p}$, and $\sigma_{\rm T}$
are the light velocity,
the black-hole mass,
the proton mass, and the Thomson scattering cross-section,
respectively.
Since the mass-accretion rate is determined 
by some mechanism which is independent of the nature of
the central star itself (Alme \& Wilson 1976; 
Spruit \& Ritter 1983; 
Norman \& Scoville 1988;
King et al. 1997; 
Umemura, Fukue, \& Mineshige 
1998; 
Ohsuga et al. 1999),
there are no reasons why $\dot{M}$ should always be below $\dot{M}_{\rm crit}$.
That is, 
the supercritical accretion flow may naturally arise.
Indeed, 
several possible sites of 
super- or near-critical accretion flow have been suggested;
e.g.,
micro-quasars
(Yamaoka, Ueda, \& Inoue 2001),
ultraluminous X-ray sources (ULXs)
(Okada et al. 1998;
Colbert \& Mushotzky 1999; 
Mizuno et al. 1999; 
Makishima et al. 2000;
Watarai et al. 2000;
Revnivtsev et al. 2002),
narrow-line Seyfert 1 galaxies
(Boller 2000; Mineshige et al. 2000),
quasars 
(Collin et al. 2002),
and Gamma-ray bursts
(Narayan, Piran, \& Kumar 2001;
Kohri \& Mineshige 2002).

When the mass-accretion rate exceeds the critical value,
photon trapping plays an important role
within the trapping radius, 
\begin{equation}
  r_{\rm trap} \sim \dot{m} \left( \frac{H}{r}\right) r_{\rm g},
  \label{rtrap}
\end{equation}
where 
$\dot{m}$ is the mass-accretion rate normalized by 
the critical mass-accretion rate, 
$H$ is the disk-half thickness,
$r$ is the radius,
and $r_{\rm g}$ is the Schwarzschild radius
(Ohsuga et al. 2002, hereafter referred to as Paper I,
see also Begelman 1978 for the spherical case).
Thus, the radiation energy, which is generated by viscosity 
in the vicinity of the black hole,
is trapped in the flow and advected inward
without being immediately radiated away. 
Since the advective energy transport 
was firstly considered 
in the slim-disk model by Abramowicz et al. (1988),
in the context of the accretion-disk models
this model has been believed to be 
the solution of the supercritical disk accretion flow
(Kato, Fukue, \& Mineshige 1998).
However, it is shown in Paper I that 
the photon-trapping effects are not accurately 
treated in the slim-disk approach.
The reason for this is that 
the time delay between the energy generation 
deep inside the disk
and the energy release at the surface
is not taken into account 
in the slim-disk formulation,
although the time delay is not negligible and 
exceeds the accretion timescale within the trapping radius.

In the light of the recent discovery 
of the supercritically accreting objects,
it is of great importance 
to study the spectral energy distribution (SED) 
of the supercritical disk accretion flow.
But, little is known about the SED
of the flow suffering the photon-trapping effects
except for the case of the slim-disk model.
Szuszkiewicz, Malkan, \& Abramowicz (1996) and Watarai et al. (2000) 
have calculated the SED of the slim disk
by superposition of blackbody/modified-blackbody spectra.
Wang et al. (1999) have solved
the radiation transfer equations in the vertical direction,
coupling with the slim-disk formulation.
As we noted above, however,
the slim-disk model has not fully considered
the photon-trapping effects, 
so we need further studies.

In paper I, we investigated 
the photon-trapping effects in the disk accretion flow,
but only solved the monochromatic radiation transfer equations
inside the accretion flow.
In this work, 
we investigate the SED of the optically-thick supercritical accretion flows
by solving multi-frequency radiation transfer equations.
In \S 2, we give our
model and the basic equations.
The resultant SEDs are presented in \S 3.
Comparing our results with the observational data, 
we discuss the role of the photon trapping
in the actual objects.
Finally, \S 4 and \S 5 are devoted to discussion and conclusions, 
respectively.

\section{OUR MODEL AND BASIC EQUATIONS}
In this section, 
we numerically solve the multi-frequency radiation-transfer equations,
as well as the energy equations of gas and radiation,
by taking into account the photon-trapping effects.
Since we are concerned with the radiation processes, 
especially the photon-trapping effects themselves,
we employ a simple model for the accretion flow structure
(see Paper I).

Here, the cylindrical coordinates, $(r, \varphi, z)$, are used.
We consider that the accretion disk is axisymmetric 
and steady in the Eulerian description; 
$\partial/\partial \varphi = \partial/\partial t = 0$.
We employ the Gaussian density profile
in the vertical direction,
\begin{equation}
  \rho(r, z) = \rho_0(r)\exp \left[-\left(\frac{z}{H}\right)^2 \right],
  \label{rho}	
\end{equation}
where $\rho_0$ is the gas density on the equatorial plane 
and we assume the ratio of the disk-half thickness to the radius, 
$H/r$, to be a constant in radius, depending on $\dot{m}$;
\begin{equation}
  \frac{H}{r} = \varepsilon_{100} \left( \frac{\dot{m}}{10^2} \right),
  \label{H}	
\end{equation}
with $\varepsilon_{100}$ being a dimensionless parameter
representing the aspect ratio, $H/r$, at $\dot{m}=100$.
Such a linear relation between $H/r$ and $\dot{m}$
exactly holds in the radiation pressure-dominated part of 
the standard-disk model
[see, e.g., equation (3.61) of Kato, Fukue, \& Mineshige (1998)],
and also roughly holds in the slim-disk model
as long as $\dot{m}$ is less than a few hundreds
(Watarai et al. 2000).
The density on the equatorial plane, 
$\rho_0$, is related to the surface density as,
\begin{equation}
  \Sigma(r) = 2 \rho_0(r) \int_{-\infty}^{\infty}
  \exp \left[-\left(\frac{z}{H}\right)^2 \right] dz
  = 2 \sqrt{\pi} \rho_0(r)H(r),
  \label{Sigma}	
\end{equation}
where the surface density, $\Sigma$, is calculated by the continuity
equation for a given accretion rate, $\dot{M}$,
\begin{equation}
  \dot{M}= -2\pi r v_r \Sigma,
  \label{mdot}	
\end{equation}
with $v_r$ being the radial component of the velocity.

Since 
the radial component of the velocity, $v_r$,
is related to the viscosity parameter $\alpha$ 
through $v_r \sim -\alpha \left( H/r \right) c_{\rm s}$
with $c_{\rm s}$ being the sound velocity
(see, e.g., Frank, King, \& Raine 1985;
Kato, Fukue, \& Mineshige 1998),
and $H$ is given by the equation (\ref{H}),
$v_r$ is finally
expressed in terms of the free-fall velocity as
\begin{equation}
   v_r = - \alpha \varepsilon_{100}^2 \left( \frac{\dot{m}}{10^2} \right)^2 
   \left( \frac{GM}{r} \right)^{1/2}.
  \label{vr}	
\end{equation}
The vertical component is prescribed as
\begin{equation}
   v_z = \frac{z}{r} v_r,
  \label{vz}	
\end{equation}
i.e., we assume a convergence flow.

Through equations (\ref{H}), (\ref{mdot}), and (\ref{vr}),
$\Sigma$ and $\rho$ are related to
$\varepsilon_{100}$ and $\dot{m}$ in our model;
$\Sigma \propto \varepsilon_{100}^{-2} \dot{m}^{-1}$
and $\rho \propto \varepsilon_{100}^{-3} \dot{m}^{-2}$.
The photon-trapping radius is rewritten 
by using equation (\ref{H}) as,
\begin{equation}
  r_{\rm trap} \sim 10^{-2} \varepsilon_{100} \dot{m}^{2} r_{\rm g}.
\label{degree}
\end{equation}
Here, it is noted that the photon-trapping radius not only indicates
the critical radius for the photon trapping
but also expresses the significance of the photon-trapping effects,
since this relation is derived from the 
comparison between the accretion timescale
and the radiative diffusion timescale.
Thus, the photon-trapping effects are more pronounced
in the case of larger $\varepsilon_{100}$ and $\dot{m}$.


We suppose the structure of the accretion disk to be 
locally plane parallel;
that is,
the disk is regarded as consisting of accreting ring elements
(see discussion in the end of this section).
Each ring is composed of $N$ layers in the $z$-direction,
and the thickness of each layer is $\Delta z = 2H_0/N$
(see Figure 2 of Paper I).
Here, $H_0$ defines the upper boundary of the disk in the
numerical simulations in such a way that $\rho(r,H_0) = 0.1 \rho_0(r)$.
We solve the time-dependent energy fields of gas and radiation
in the ring element during the course of accretion motion
until the element reaches the inner edge of the disk.
Both of the viscous heating and compressional heating
though converging inflow are considered.
Since the flow assumed to be steady ($\partial/\partial t = 0$),
time coordinate can be transformed to the spatial coordinates;
i.e., $D/Dt=v_r\partial/\partial r+v_z\partial/\partial z$
with $v_r$ and $v_z$ being given by equations (\ref{vr}) and (\ref{vz}).
We thus express the multi-frequency radiative flux at the disk surface 
as a function of radius 
and then calculate the resultant SED and luminosity of the disk.

In the plane-parallel approximation,
the radial and azimuthal components of the radiative flux 
are null.
Also, non-diagonal components of the radiation stress tensor
are null.
Hence, using equations (\ref{vr}) and (\ref{vz}),
we write the energy equations of radiation and gas as 
\begin{eqnarray}
   \rho \left( v_r \frac{\partial}{\partial r}
   +\frac{z}{r} v_r \frac{\partial}{\partial z} \right)
   \left(\frac{E_\nu}{\rho}\right) \nonumber\\
   = -\frac{\partial F_{\nu}^z}{\partial z}
   -\frac{v_r}{r} 
   \nu \frac{\partial}{\partial \nu}	
   && \left( \frac{1}{2} P_{\nu}^{rr}
   -P_{\nu}^{\varphi\varphi}
   -P_{\nu}^{zz}
   \right) \nonumber\\
   +4\pi \kappa_{\nu} B_{\nu} - c\kappa_{\nu}E_{\nu},&&
\label{radene1}
\end{eqnarray}
and 
\begin{eqnarray}
   \rho \left( v_r \frac{\partial}{\partial r}
   +\frac{z}{r} v_r \frac{\partial}{\partial z} \right)
   \left(\frac{e}{\rho}\right) \nonumber\\
   = -\frac{3}{2}\frac{v_r}{r}p_{\rm gas}  
   -4\pi \kappa_P B & & + c\kappa_E E
   + q_{\rm vis},
\label{gasene}
\end{eqnarray}
respectively,
where 
$E_\nu$ and $E (\equiv \int E_\nu d\nu)$ are the radiation energy density and 
its frequency-integrated form, respectively,
$F_\nu^z$ is the radiative flux in the $z$-direction,
$P_\nu^{ii}$ is the diagonal ($ii$-th) 
components of the radiation stress tensor,
$B_\nu$ and $B=\sigma T_{\rm gas}^4/\pi$ are the Planck function and
its frequency-integrated form, respectively,
with $T_{\rm gas}$ being the gas temperature
and $\sigma$ being the Stefan-Boltzmann constant,
$\kappa_\nu= 
3.7\times 10^8 T_{\rm gas}^{-1/2} (\rho/m_{\rm p})^2 \nu^{-3}
\{1-\exp(-h\nu/kT_{\rm gas})\}$ 
is the free-free absorption coefficient
(Rybicki \& Lightman 1979),
with $h$ being the Planck constant and 
$k$ being the Boltzmann constant,
$e$ is the internal energy density,
$p_{\rm gas}$ is the gas pressure,
$\kappa_P$ and $\kappa_E$ are the Planck-mean and energy-mean opacities,
respectively,
\begin{equation}
   \kappa_P = \frac{1}{B}\int \kappa_\nu B_\nu d\nu,
   \label{kappaP}
\end{equation}
\begin{equation}
   \kappa_E = \frac{1}{E}\int \kappa_\nu E_\nu d\nu,
   \label{kappaE}
\end{equation}
and $q_{\rm vis}$ is the viscous heating rate per unit volume
(Mihalas \& Klein 1982; Mihalas \& Mihalas 1984; 
Fukue, Kato, \& Matsumoto 1985;
Stone, Mihalas, \& Norman 1992).

To close the set of equations,
we apply the multi-frequency flux-limited diffusion approximation (FLD),
for the radiative flux and stress tensor,
\begin{equation}
   F_\nu^z = - \frac{c\lambda_\nu}{\chi_\nu}\frac{\partial E_\nu}{\partial z},
\end{equation}
\begin{equation}
   P_\nu^{rr} = P_\nu^{\varphi\varphi} = \frac{1}{2}(1-f_\nu)E_\nu,
\end{equation}
and 
\begin{equation}
   P_\nu^{zz} = f_\nu E_\nu,
\end{equation}
with $\chi_\nu$ being the extinction coefficient,
\begin{equation}
   \chi_\nu=\frac{\rho\sigma_{\rm T}}{m_{\rm p}}+\kappa_\nu,
\end{equation}
where $\lambda_\nu$, $R_\nu$, and $f_\nu$ are defined as 
\begin{equation}
   \lambda_\nu = \frac{2+R_\nu}{6+3R_\nu+R_\nu^2},
\end{equation}
\begin{equation}
   R_\nu = \frac{1}{\chi_\nu E_\nu} 
   \left| \frac{\partial E_\nu}{\partial z} \right |,
\end{equation}
and 
\begin{equation}
   f_\nu=\lambda_\nu + \lambda_\nu^2 R_\nu^2.
\end{equation}
These relations correspond to the multi-frequency form
of the monochromatic FLD represented by Turner \& Stone (2001).
This approximation holds
both in the optically thick and thin regimes.
In the optically thick limit ($R_\nu \rightarrow 0$), we find
$\lambda_\nu \rightarrow 1/3$ and $f_\nu \rightarrow 1/3$.
In the optically thin limit of $R_\nu \rightarrow \infty$,
on the other hand, 
we have $| F_\nu^z | = cE_\nu$, $P_\nu^{rr}=P_\nu^{\varphi\varphi}=0$,
and $P_\nu^{zz}=E_\nu$.
These give correct relations in the optically thick diffusion limit
and optically thin streaming limit, respectively.
Thus, we can rewrite the radiation energy equation (\ref{radene1}) as
\begin{eqnarray}
   \rho \left( v_r \frac{\partial}{\partial r}
   +\frac{z}{r} v_r \frac{\partial}{\partial z} \right)
   \left(\frac{E_\nu}{\rho}\right) \nonumber\\
   = c\frac{\partial}{\partial z} \left(
   \frac{\lambda_\nu}{\chi_\nu}\frac{\partial E_\nu}{\partial z} \right)
   +\frac{v_r}{r}&\nu&\frac{\partial}{\partial\nu}
   \left( \frac{3f_\nu+1}{4}E_\nu \right) \nonumber\\
   +4\pi \kappa_\nu B_\nu - c\kappa_\nu E_\nu.&&
\label{radene2}
\end{eqnarray}

These nonlinear equations (\ref{gasene}) and (\ref{radene2}) are integrated 
iteratively by the Newton-Raphson method 
with the Gauss-Jordan  elimination for a matrix inversion,
by being coupled with 
the equation of state,
\begin{equation}
   p_{\rm gas} = \frac{2}{3}e.
\end{equation}
We employ the simple model for viscous heating rate,
whose distribution in the $z$-direction 
is proportional to the total pressure;
that is,
\begin{equation}
   q_{\rm vis}(r,z) = Q_{\rm vis}(r)\frac{E(r,z)+2 e(r,z)/3}
   {\int_0^{H_0} \left\{ E(r,z)+2 e(r,z)/3 \right\} dz},
   \label{qvis}
\end{equation}
where $Q_{\rm vis}$ is the half 
vertically-integrated viscous heating rate
per unit surface,
\begin{equation}	
  Q_{\rm vis} \sim \frac{3}{8\pi} \Omega_{\rm K}^2 \dot{M}
  \left[ 1-\left( \frac{r_{\rm in}}{r} \right)^{1/2} \right],
\label{Qvis}
\end{equation}
(Shakura \& Sunyaev 1973; Lynden-Bell \& Pringle 1974),
with $\Omega_{\rm K}$ being the Keplerian angular speed.

Throughout the present study, $\alpha$ and $N$ are set
to be 0.1 and 100, respectively.
We consider the range of frequency, $\log \nu = 16-20$, 
with $\Delta \log\nu = 0.08$.
Since the radiative flux from the vicinity of the black hole
gives negligible contribution to the total SED
due to the large redshift in the face-on case, 
we neglect the emission from the regions inside, $r<3r_{\rm g}$
[see, e.g., equation (3.159) of Kato, Fukue, \& Mineshige (1998)].
The supercritical disk becomes relatively geometrically thick through
the strong radiation pressure and is obscured by the outer part of
the disk in the edge-on case so that it would be observed 
as objects of small inclination angle.
The black-hole mass is fixed to be $10 M_\odot$ except
in \S 3.4, where we adopt $50 M_\odot$ and $100 M_\odot$ 
to compare our results with 
the observations of ULXs.

In our simulations, 
we assumed the disk to be locally plane parallel.
Hence, the radiation energy diffuses towards only the $z$-direction
in the frame moving with gas;
i.e., the radial and azimuthal components of the radiative flux 
are null.
However, if the radial gradient of the radiation energy density
is comparable to the vertical one,
$|\partial E/\partial z| \sim |\partial E/\partial r|$,
the diffusion of the radiation energy towards the $r$-direction
is not negligible.
Such a situation would hold in the very vicinity of 
the inner edge of the disk,
where the radiation energy effectively falls onto the
black hole because of $F^r < 0$.
Thus, the photon trapping would be enhanced at $r\lsim 3r_{\rm g}$.
In contrast, 
the radiative flux would become positive, $F^r > 0$,
at $r > 3r_{\rm g}$, and may not be negligible
if the aspect ratio, $H/r$, is of the order of unity.
In this case, the photon-trapping effects would be reduced.
Such a reduction of the photon trapping is roughly estimated as follows.
Since the radiative diffusion timescale is given by $c/\tau$,
the ratio of the radial diffusion velocity and 
vertical diffusion velocity
is roughly obtained to be $v_{\rm diff,r}/v_{\rm diff,z}
\sim \tau_z/\tau_r \sim H/r$.
Here, $\tau_z \sim \rho \sigma_{\rm T} H/m_{\rm p}$ and 
$\tau_r \sim \rho \sigma_{\rm T} r/m_{\rm p}$ are the optical depth 
for the Thomson cross-section
in vertical and radial directions, respectively. 
By taking into account the radiative diffusion in the radial direction,
the timescale for the radiation energy to be swallowed by the black
hole is $\tau_{\rm swa} \sim r/(v_r-v_{\rm diff, r})$.
By comparison between this timescale and
the diffusion timescale in vertical direction, $H/(c/\tau_z)$,
we find that the photon-trapping radius is modified as
\begin{equation}
   r'_{\rm trap} \sim r_{\rm trap}
   \left\{ 1+\left(\frac{H}{r} \right)^2 \right\}^{-1},
\end{equation}
where equations (\ref{rtrap}), (\ref{mdot}), 
$\tau_z = \sigma_{\rm T}\Sigma/m_{\rm p}$,
$\dot{M}_{\rm crit} \equiv L_{\rm E}/c^2$,
and
$L_{\rm E}=4\pi cGMm_{\rm p}/\sigma_{\rm T}$
are used.
Since the luminosity depends on the trapping radius as 
$L \sim \int_{r_{\rm trap}} 2\pi r Q_{\rm vis} dr \propto r_{\rm trap}^{-1}$ 
(Paper I),
the luminosity would increase by a factor of $1+(H/r)^2$
on account of the radiative diffusion in the radial direction.
For $\varepsilon_{100}=0.5$, 
our formulation underestimate luminosity by 
$(H/r)^2 \sim 0.25 (\dot{m}/100)^2$.
Doppler beaming tends to focus radiation field 
to be more confined within the equatorial plane, 
thereby enhancing the photon trapping effects.
The details would be elucidated
by solving two-dimensional radiation transfer equations
as future work.

\section{RESULTS}
\subsection{Luminosity}
We first plot the luminosity against the mass-accretion rate in Figure 1,
where the luminosity and mass-accretion rate are 
normalized by the Eddington luminosity and the critical mass-accretion rate,
respectively.
The luminosity should increase along the dashed line,
if the energy-conversion efficiency, $\eta\equiv \dot{M}c^2/L$, is constant.
However, it is found that 
the energy-conversion efficiency decreases
with increase in the mass-accretion rate
at $L \gsim 2L_{\rm E}$ due to the photon trapping.
We also see that the larger $\varepsilon_{100}$ is,
the lower becomes the luminosity at high
mass-accretion rates, $\dot{m} \gg 10$.
This can be understood,
since the photon-trapping effects are more conspicuous 
for larger $\varepsilon_{100}$
[see equation (\ref{degree})].
We have confirmed the results of Paper I,
whereby energy-conversion efficiency decreases
with increase of the mass-accretion rate,
by the multi-frequency radiation transfer simulations.
\centerline{\psfig{file=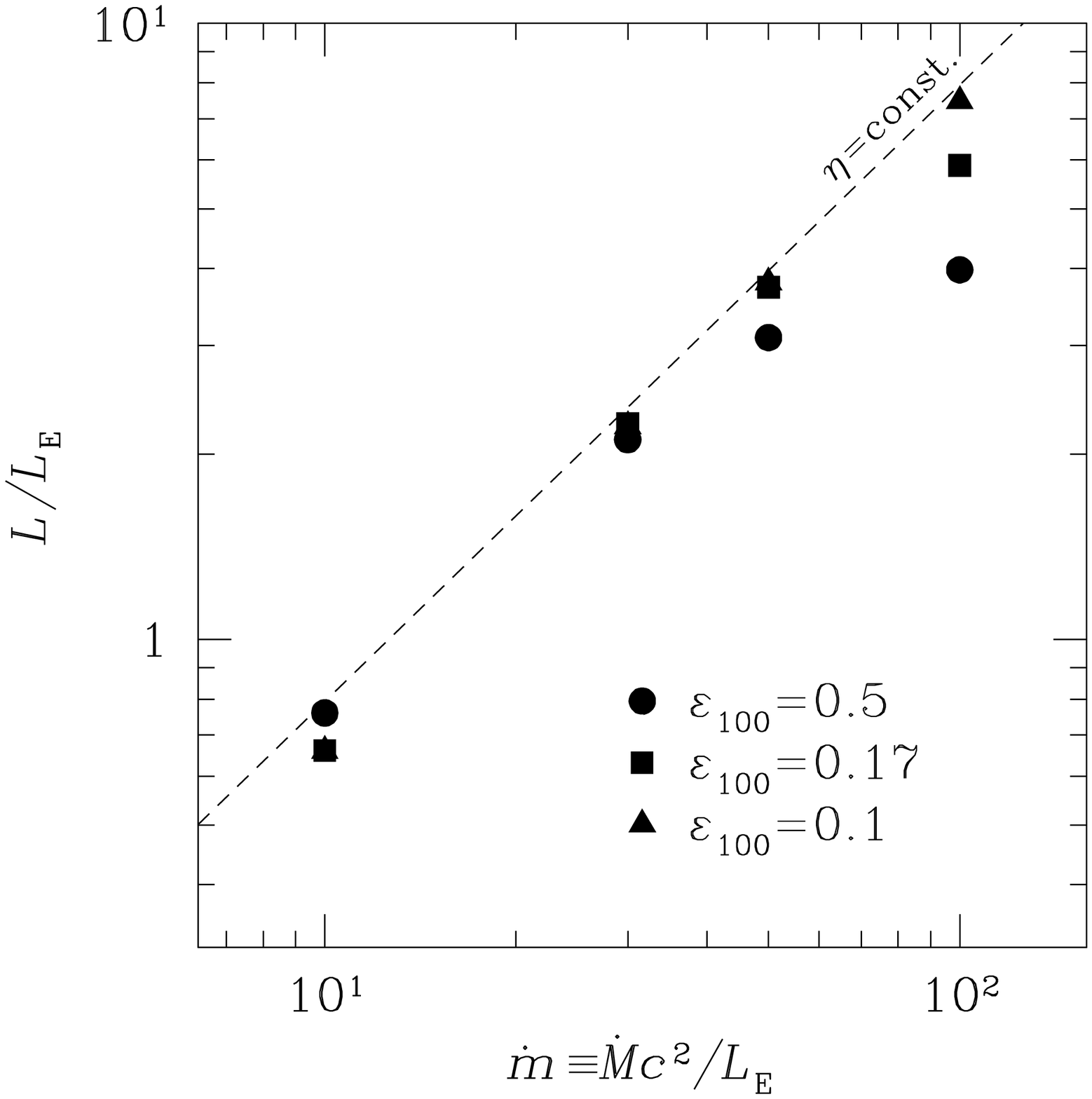,width=3.4in}}
\figcaption[fig1.eps] {
Total disk luminosity as a function of the mass-accretion rate,
$\dot{m}\equiv \dot{M}c^2/L_{\rm E}$.
The circles, squares, and triangles represent the cases with
$\varepsilon_{100}=0.5$, 0.17, and 0.1, respectively,
where $\varepsilon_{100}$ is a dimensionless parameter
defined as $\varepsilon_{100}\equiv (H/r)/(\dot{m}/100)$.
The dashed line shows the luminosity 
for the cases that the energy-conversion efficiency, 
$\eta\equiv \dot{M}c^2/L$, is kept constant.
\label{fig1}}

\subsection{Effective Temperature}
The effective temperature profile is shown in Figure 2.
Here, the effective temperature is defined as 
$T_{\rm eff}=(F/\sigma)^{1/4}$, where
$F=\int F_\nu d\nu$.
The profile becomes a bit flatter in the vicinity of the 
black hole when $\dot{m} \gsim 30$
due to the photon-trapping effects.
In particular, the slope becomes somewhat flatter than
that predicted by the slim-disk model, $r^{-1/2}$,
in the case of $\dot{m}=100$.
When $\dot{m}=10$, in contrast, 
the profile is consistent with the standard-disk model, 
in which $T_{\rm eff} \propto r^{-3/4}$,
except at $r/r_{\rm g} \lsim 7$,
where boundary effects appear,
since the photon-trapping effects are negligible.
Here, it should be stressed that 
the flatter effective temperature means 
a reduction of the radiative flux at small radii,
however, 
the local emergent spectrum is no longer 
a simple multi-color blackbody spectrum
(discussed in the next subsection).
\centerline{\psfig{file=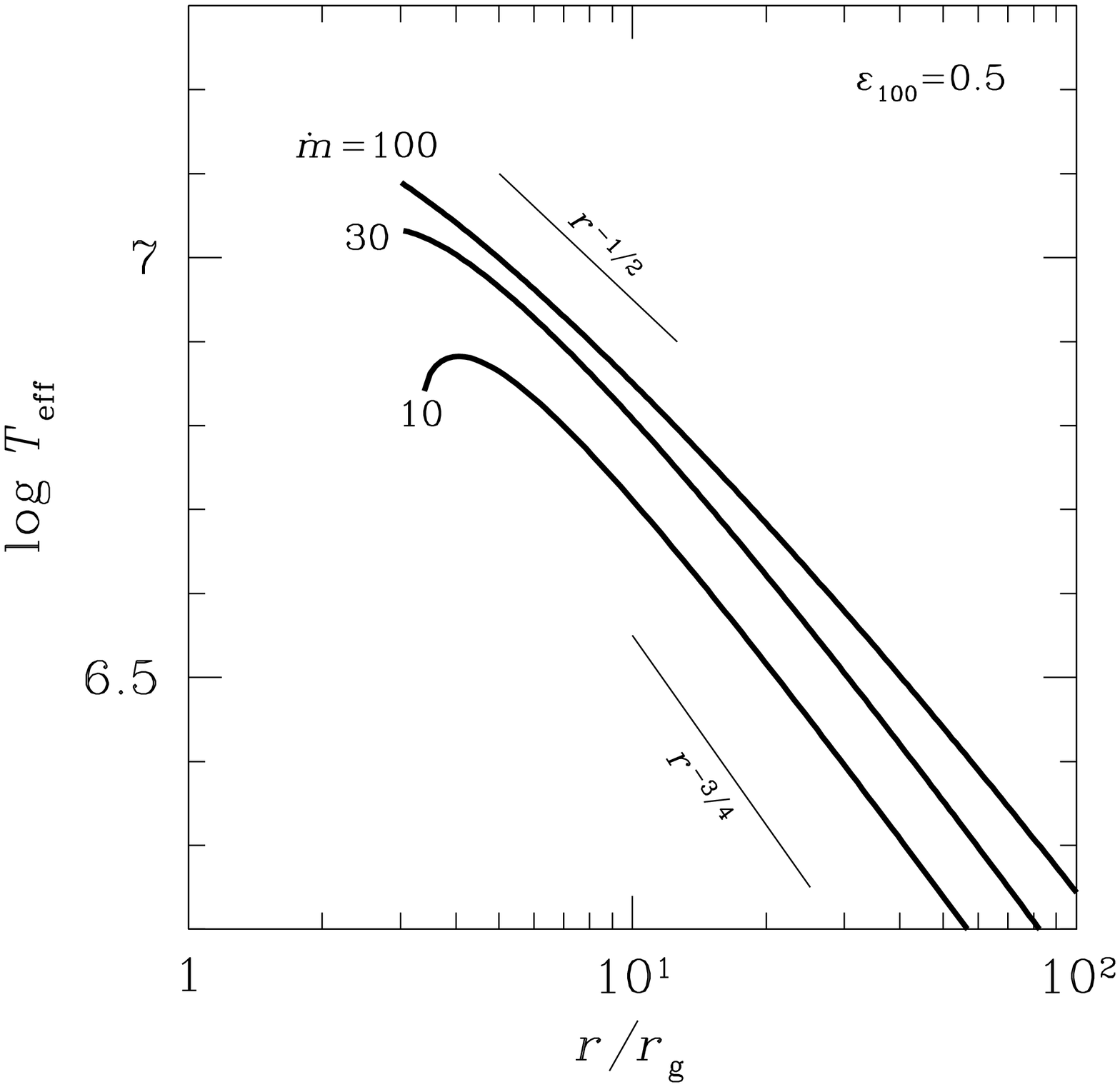,width=3.4in}}
\figcaption[fig2.eps] {
The effective temperature profiles for the case with $\varepsilon_{100}=0.5$
for various $\dot{m}$; $\dot{m}=10$, 30, and 100.
Through the photon-trapping effects,
the profile becomes flatter in the vicinity of the black hole
as $\dot{m}$ increases beyond $\dot{m} \sim 30$.
In particular, the profile for the case with $\dot{m}=100$ 
is somewhat flatter than 
the prediction of the slim-disk model; i.e., 
$T_{\rm eff} \propto r^{-1/2}$.
\label{fig2}}

\subsection{Spectral Energy Distribution}
We show the SED of the accretion flow with $\dot{m}=10$
for $\varepsilon_{100}=0.5$, 0.17, and 0.1 in Figure 3.
For such a low $\dot{m}$, the photon-trapping effects do not appear
yet,
but the calculated SEDs grossly deviate from the SED 
predicted by the standard disk with the same $\dot{m}$, which is indicated
by the dotted curve.
It is found that the SED is shifted towards the higher-energy side
as $\varepsilon_{100}$ increases.
This is explained by the transmission of high-energy photons generated
near the equatorial plane,
since for given $\dot{m}$ both of $\Sigma$ and $\rho$ decrease
as $\varepsilon_{100}$ increases,
leading to a decrease of $\kappa_\nu$, especially at high $\nu$
(note $\kappa_\nu \propto \rho^2$ and $\kappa_\nu$ 
decreases as $\nu$ increases).
Then, the high-energy photons arising from deep inside the disk
can pass through the disk body when $\varepsilon_{100}$ is large.
In the case of $\varepsilon_{100}=0.5$ and 0.17, 
the frequency of the SED peak intensity (hereafter referred to as the 
peak frequency)
is by about three times higher than that of the standard disk.
This result might resolve the so-called
\lq too hot accretion disk problem' (Makishima et al. 2000).
That is, the black-hole masses of some ULXs are inferred to be 
at least $\sim 100M_\odot$,
since their luminosities reach $\sim 10^{40} {\rm erg \ s^{-1}}$,
while the source luminosities can not exceed
the Eddington luminosity.
Then, the standard-disk model predicts that 
the innermost disk temperatures 
of these objects 
should be $kT_{\rm in} <$ 1 keV for $M \gsim 100 M_\odot$,
however, the observed temperatures are much higher (1.1-1.8 keV).
This is the \lq too hot accretion disk problem.'
\centerline{\psfig{file=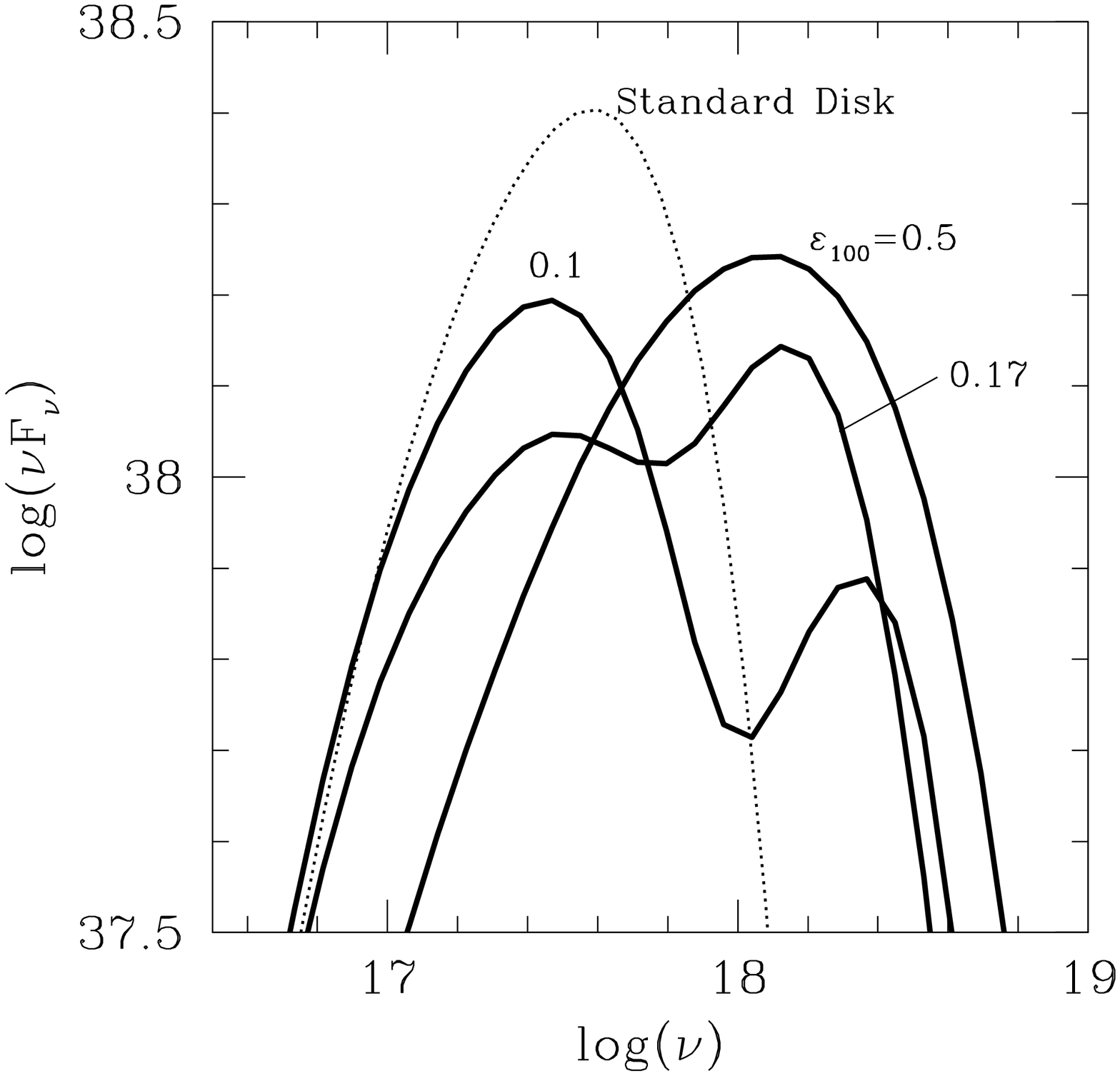,width=3.4in}}
\figcaption[fig3.eps] {
The spectral energy distribution (SED) of the accretion disk for $\dot{m}=10$
and for $\varepsilon_{100}=0.5$, 0.17, and 0.1.
The dotted curve shows the SED predicted by the standard-disk model.
The SED is shifted towards the higher-energy side with the 
increase of $\varepsilon_{100}$,
since for large $\varepsilon_{100}, \Sigma$, $\rho$, 
and, hence, $\kappa_\nu$ are smaller so that
the high-energy photons arising from deep inside the disk
can easily pass through the disk body.
\label{fig3}}

\vspace{2mm}
In our simulations,
the spectral hardening of the emergent SED is caused by 
the transmission of the high-energy photons 
arising from deep inside the disk.
The Comptonization is not taken into account
in our simulations 
although it provides another important cause of the hardening.
The hardening due to the Comptonization
was demonstrated by Czerny \& Elvis (1987) and
Ross, Fabian, \& Mineshige (1992)
[see, however, Laor \& Netzer (1989) for negative results].
Shimura \& Takahara (1993) solved
multi-frequency radiation transfer equations coupled with
the hydrostatic balance and Comptonization,
finding the spectral hardening factor of $\kappa \sim 1.7$.

When $\varepsilon_{100}$ is small, say $\varepsilon_{100}=0.1$,
$\Sigma$ and $\kappa_\nu$ are large, and, therefore, 
photons tend to be absorbed and re-emitted before reaching
the disk surface.
Thus, only low-energy photons which are generated near the disk surface
can go out and thus the emergent SED 
becomes similar to that of the standard disk.
However, 
since the absorption coefficient by the free-free absorption 
strongly depends on the frequency as $\nu^{-3}$,
a part of high-energy photons can directly pass through the disk,
and a double peak appears 
(see the cases with $\varepsilon_{100}=0.17$ and 0.1).
Here, we should note that the higher-energy peak might disappear,
if the metal absorption,
which can be much more effective at high frequency,
is included (discussed later).
	
The emergent SEDs in the case of $\dot{m}=100$ is 
shown in Figure 4.
As was mentioned above,
the peak frequency 
tends to increase as $\varepsilon_{100}$ increases
due to the transmission of the high-energy photons.
However, the SED in the case of $\varepsilon_{100}=0.5$
is softer than other cases.
This is caused by the photon-trapping effects.
Since the higher-energy photons 
originate from deep inside the disk,
they tend to be more effectively trapped in the flow
than lower-energy ones.
As a consequence,
the peak frequency of the emergent SED 
eventually turns to decrease as $\dot{m}$ further increases.
That is, the photon trapping provides the counter effect to 
the transmission of the high-energy photons.
Thus, the peak frequency of emergent photons 
starts to decrease when $\dot{m}$ exceeds some critical value
[see equation (\ref{degree})].
\centerline{\psfig{file=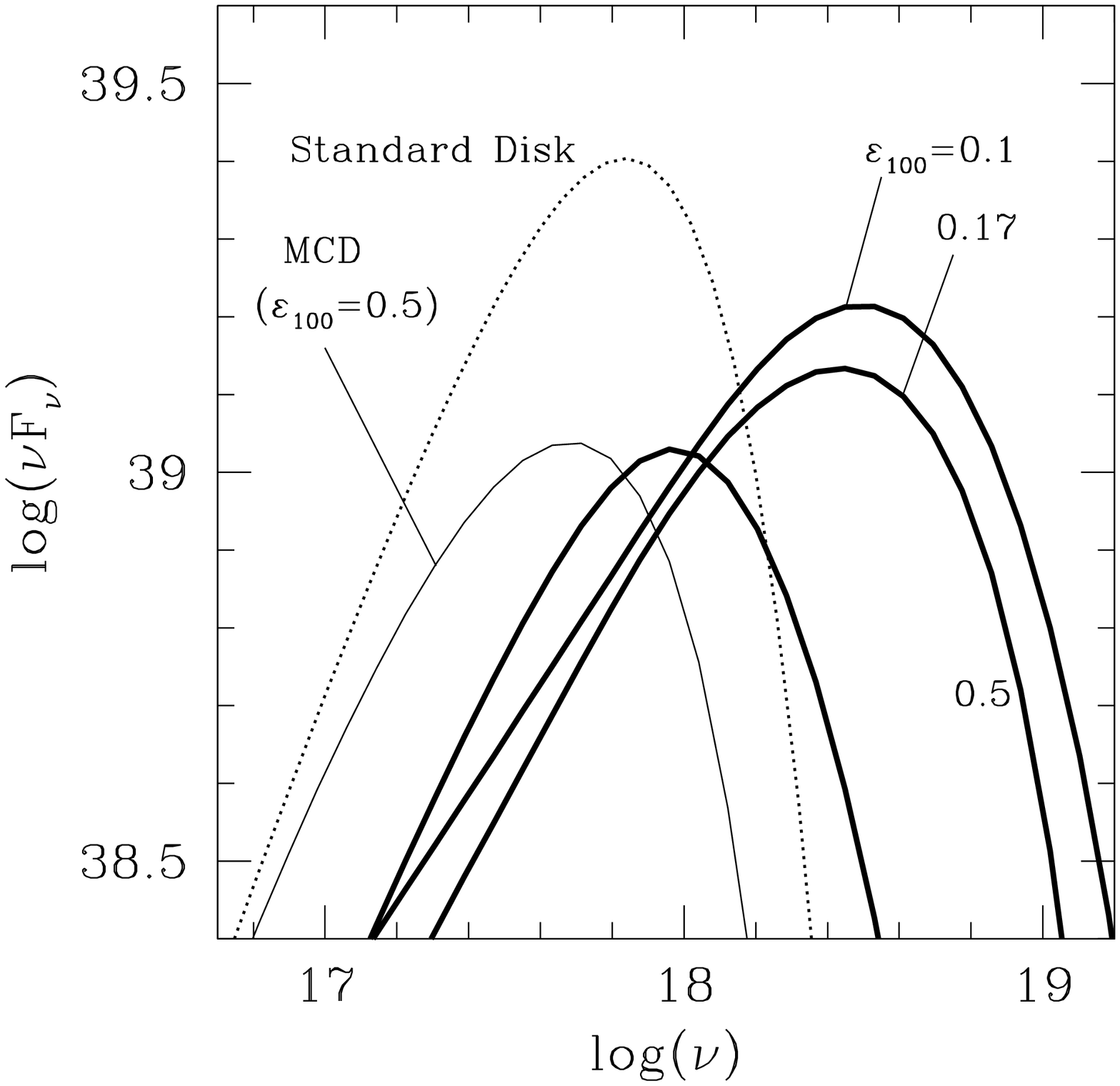,width=3.4in}}
\figcaption[fig4.eps] {
Same as Figure 3, but for $\dot{m}=100$.
The thin solid curve indicates the SED of the multi-color disk (MCD) model.
The peak frequency of the SED becomes considerably low
in the case of $\varepsilon_{100}=0.5$ due to effective photon trapping.
The MCD model does not correctly reproduce the actual SED.
\label{fig4}}

\vspace{2mm}
In Figure 4, 
the thin solid curve is the SED of the multi-color disk (MCD) model
for $\varepsilon_{100}=0.5$,
where the SED of the MCD model is calculated by the superposition
of the blackbody spectra with various effective temperature, 
$B_\nu (T_{\rm eff})$ with $T_{\rm eff} \equiv (F/\sigma)^{1/4}$.
As shown in this figure, 
the emergent SED is deviated from that of the MCD model.
Therefore, 
not only the standard-disk model but also 
the MCD model can not successfully reproduce the actual SEDs
in the supercritical accretion flow,
and it is necessary to incorporate photon-trapping effects
accurately when $\dot{m}\gsim 100$.

In Figure 5, we plot the variations of SED 
with changes in $\dot{m}$
for $\varepsilon_{100}=0.17$ and 0.5.
Symbols filled squares and circles 
indicate the locations of the SED peak.
The peak frequency of the SED increases
as the mass-accretion rate increases up to $\dot{m}=30$
for both cases.
When $\dot{m}$ is larger than 30,
however, the peak frequency no longer increases,
since the photon trapping works and tends to suppress 
the emergence of high-energy photons.
In the case of $\varepsilon_{100}=0.5$,
in which the photon-trapping effects are more enhanced
than the case of $\varepsilon_{100}=0.17$,
the peak frequency rather decreases with 
increase of the mass-accretion rate.
\centerline{\psfig{file=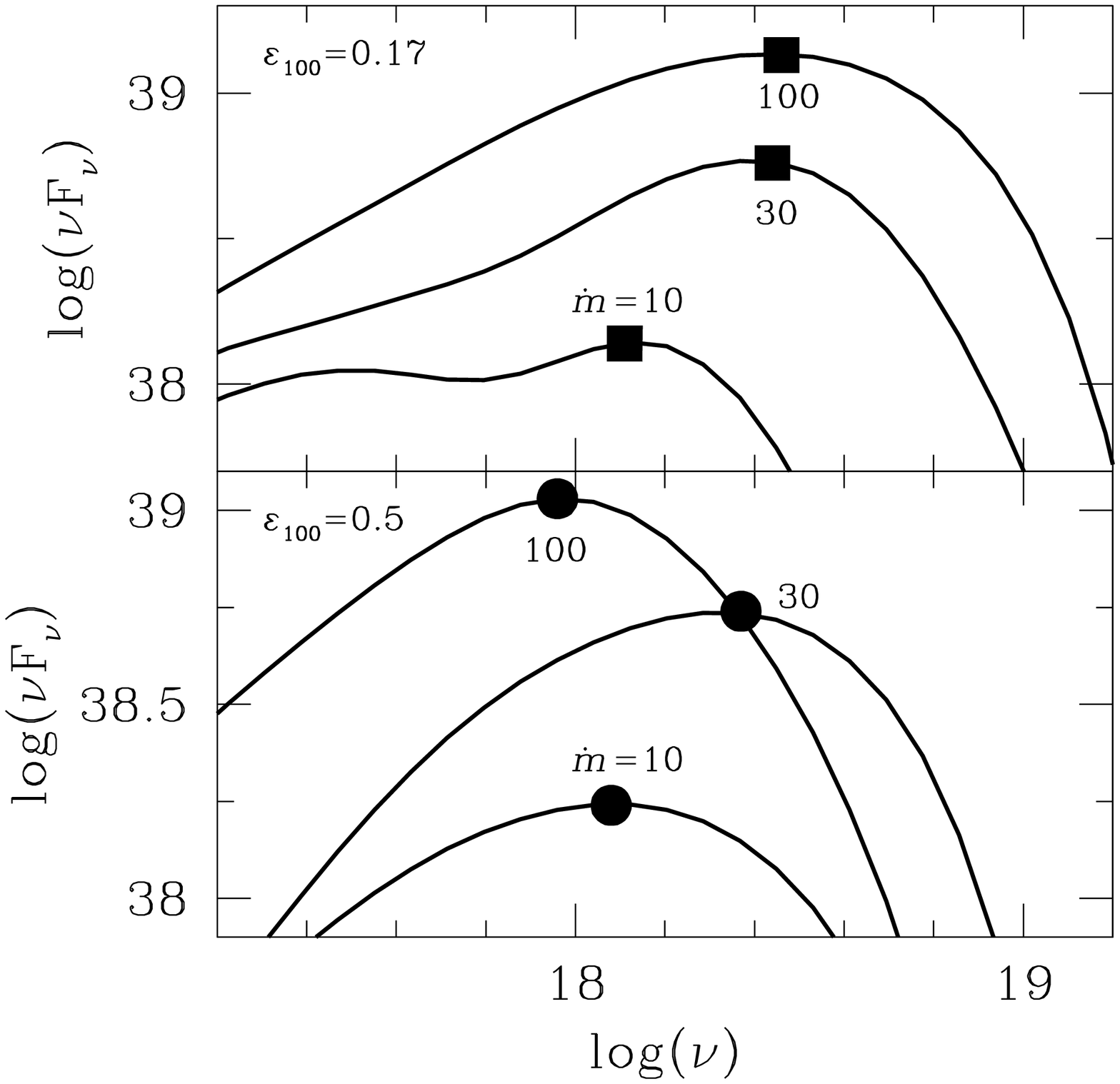,width=3.4in}}
\figcaption[fig5.eps] {
The SED for $\varepsilon_{100}=0.17$ (upper panel) and 0.5 (lower panel)
for different mass-accretion rates,
$\dot{m}=10$, 30, and 100.
In contrast with the standard disk regime at $\dot{m}\leq 30$,
in which the peak frequency increases with increasing $\dot{m}$,
the peak frequency of the SED does not increase
with an increase of mass-accretion rate
when the photon trapping is effective ($\dot{m}\geq 30$).
Rather, the peak frequency decreases
as $\dot{m}$ increase, when $\varepsilon_{100}=0.5$. 
\label{fig5}}

\vspace{2mm}
To sum up, 
the SED of the supercritical accretion flow
shifts towards the higher-energy side 
owing to the transmission of high-energy photons
as long as the photon-trapping effects are not appreciable.
When $\dot{m}$ exceeds $\sim 30$, however, 
the photon-trapping effects become substantial 
if $\varepsilon_{100} \gsim 0.17$,
and the peak frequency starts to decrease
in spite of an increase of the mass-accretion rate.

\subsection{Spectral fitting}
\subsubsection{X-ray H-R diagram}
In the previous subsection,
we overviewed the emergent SED 
in the supercritical accretion flow
with careful considerations on the photon-trapping effects.
Here, 
we fit the calculated spectra, following Watarai et al. (2000).
In this fitting method, 
a simple temperature profile, 
$T_{\rm in} (r/r_{\rm in})^{-3/4}$
with $T_{\rm in}$ and $r_{\rm in}$ being fitting parameters,
is employed for the model SED 
(Mitsuda et al. 1984).
We adopt the photon energy range between 0.1keV and 5keV for the fitting,
but we should keep in mind that
the fitting results depend on the adopted energy range,
in which the fitting is made.
We plot the calculated
$T_{\rm in}$ as a function of $L$
in comparison with the observational data
in the X-ray H-R diagram (see Figure 6).

As shown in Figure 6,
$kT_{\rm in}$ increases as the mass-accretion rate increases
for the case of $M=10M_\odot$ and $\varepsilon_{100}=0.17$
(filled squares).
In the case of $\varepsilon_{100}=0.5$ (filled circles),
in contrast,
the innermost temperature $T_{\rm in}$ decreases with 
increase of $\dot{m}$
when $\dot{m}>30$.
This is due to very efficient photon-trapping effects,
by which the emergent SED shifts towards the lower-energy side,
leading to a decrease in $kT_{\rm in}$.
As was mentioned in the previous subsection,
the photon-trapping effects are more conspicuous 
in the accretion flow with larger $\varepsilon_{100}$.
\centerline{\psfig{file=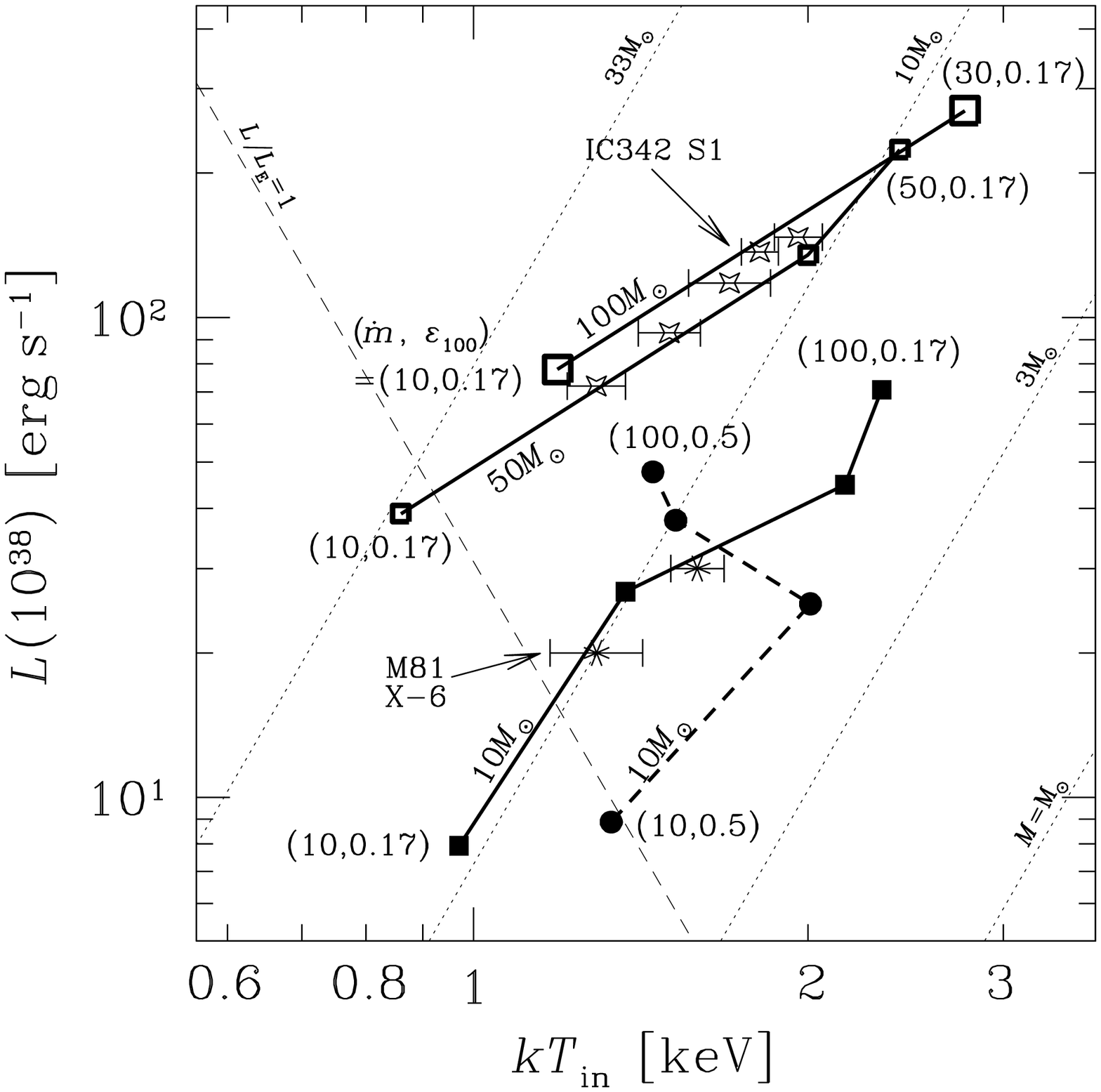,width=3.4in}}
\figcaption[fig6.eps] {
The predicted and observed trajectories of black-hole object 
with varying luminosity in the X-ray H-R diagram. 
The filled circles and filled squares indicate our numerical results
for $\varepsilon_{100}=0.5$ and 0.17, respectively, with different 
mass-accretion rates, $\dot{m}=10$, 30, 50, and 100
from the bottom to the top.
The small and big open squares represent 
the results for $\varepsilon_{100}=0.17$, 
but the black-hole mass is 
$M=50M_\odot$ and $100M_\odot$, respectively.
A result of $\dot{m}=100$ is not shown
for $M=50M_\odot$,
and we plot only results of $\dot{m}=10$ and 30
for $M= 100M_\odot$.
The dotted lines correspond to the cases of the standard-disk model
with different black-hole mass, $M=1M_\odot$, $3M_\odot$, 
$10M_\odot$, and $33M_\odot$.
According to the standard-disk model, the luminosity becomes 
the Eddington luminosity on the thin dashed line.
The case of $(M, \varepsilon_{100}) = (10M_\odot, 0.17)$
nearly coincides with the observational results 
of M81 S-6.
Moreover, the observed variation of IC 342 S1 can be 
reproduced by our model 
for $(M, \varepsilon_{100}) = (50-100M_\odot, 0.17)$.
\label{fig6}}

\vspace{2mm}
In this figure,
we found that 
the case of $\varepsilon_{100}=0.17$ (filled squares) 
coincides most with the observational results 
of M81 S-6.
The cases of $\varepsilon_{100}=0.5$ 
(filled circles)
does not reproduce the observed trend.
Therefore, we can rule out
the cases of $\varepsilon_{100}=0.5$, that is,
the photon-trapping effects should only moderately appear 
in the accretion flow of ULXs.
This conclusion is supported by the fitting to the
IC 342 S1 data, which is also well reproduced by the cases
with $\varepsilon_{100}=0.17$,
if the black-hole mass is scaled up 
to $M=50-100M_\odot$
(small and big open squares).

From the fitting to the IC 342 S1 data,
we find that the black hole in IC 342 S1 is 
considerably massive; its mass is $50-100M_\odot$.
Note that our estimation is by a factor of $1.7-3$ larger than
the value $\sim 30M_\odot$ which 
Watarai, Mizuno, \& Mineshige (2001) 
estimated based on the slim-disk model.
Scatters of IC 342 S1 data in the H-R diagram 
is understood in term of 
the variation of the mass-accretion rate according to our model.
If we adopt the standard-disk model,
the black-hole mass should change
to fit the observation 
which is, of course, unreasonable.
Our model is certainly preferable.

\subsubsection{Inner-edge radius and hardening factor}
Next, we estimate the inner-edge radius, $R_{\rm in}$, by adopting
equation (4) in Makishima et al. (2000) to our calculated SEDs.
We assume that the hardening factor is $\kappa=1.7$
and the correction factor to $R_{\rm in}$, 
which is required to remove boundary effects is 
set to be $\xi=0.42$.
As shown in the upper panel of Figure 7,
the resultant inner-edge radius 
depends on the parameter, $\varepsilon_{100}$, 
as well as
the mass-accretion rate, $\dot{m}$; 
namely it depends on 
how important the photon-trapping effects are.
The photon-trapping effects tend to increase the inner-edge radius
and decrease of $T_{\rm in}$ 
because of suppression of emergence of high-energy photons.
In the case of $\varepsilon_{100}=0.5$,  
$R_{\rm in}$ drastically shifts from 40 km to 160 km 
as the mass-accretion rate increases from $\dot{m}=30$ to 100
due to very efficient photon trapping.
If $\varepsilon_{100}=0.17$, conversely,
the radius decreases 
with increase in the mass-accretion rate as long as $\dot{m}\lsim 50$.
At higher $\dot{m}>50$, the radius slightly increases
due to the mild photon-trapping effects.
\centerline{\psfig{file=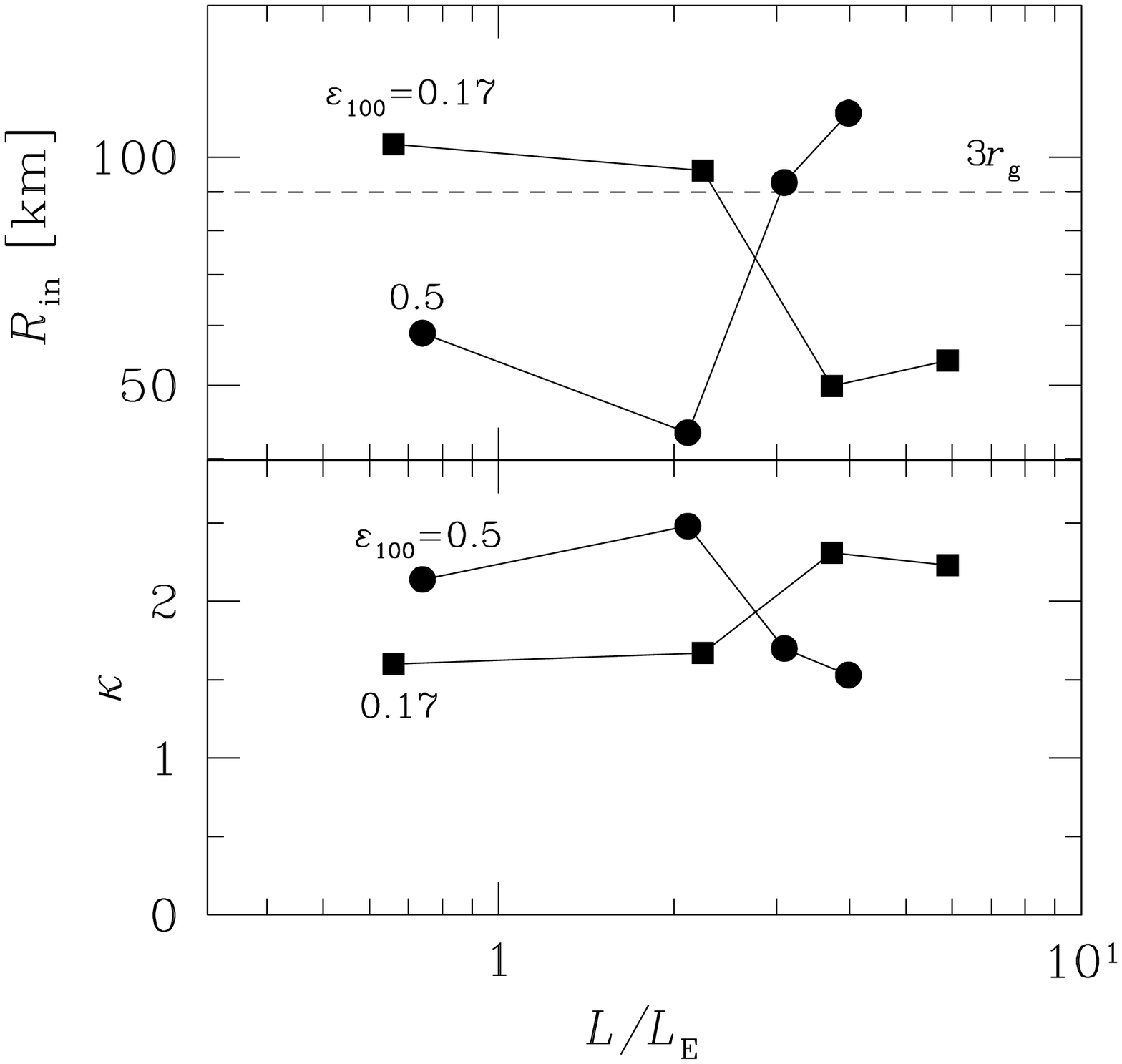,width=3.4in}}
\figcaption[fig7.eps] {
The inner-edge radius, $R_{\rm in}$ (upper panel) and 
the hardening factor,  $\kappa$ (lower panel)
for the different mass-accretion rates, 
$\dot{m}=10$, 30, 50, and 100,
from the left to the right, respectively.
\label{fig7}}

\vspace{2mm}
In the lower panel of Figure 7,
we plot the hardening factor, $\kappa$,
by employing equation (4) in Makishima et al. (2000).
Here, we set $R_{\rm in}=3r_{\rm g}$ and $\xi=0.412$.
It is clear in this figure that 
the hardening factor is reduced by the photon-trapping effects.
In the case of $\varepsilon_{100}=0.5$,
$\kappa$ decreases when $\dot{m}$ exceeds 30.
If $\varepsilon_{100}=0.17$, $\kappa$ slightly increases
for $\dot{m}<50$,
but it decreases at $\dot{m}>50$ 
when photon-trapping effects are substantial.

\section{DISCUSSION}
\subsection{Comparison with the Slim-disk Model}
In some slim-disk approach, 
the emergent SED is assumed to be 
the superposition of blackbody spectra
with different effective temperatures
(Szuszkiewicz, Malkan, \& Abramowicz 1996; 
Watarai et al. 2000).
However, such a method 
is not successful in reproducing the actual SED
as was shown in \S 3.3.
Moreover, 
we found that 
the effective temperature profiles may deviate from
the prediction of the slim-disk model, $r^{-1/2}$, at large $\dot{m}$
since the photon-trapping effects are not accurately 
treated in the slim-disk model (\S 3.2, see also Paper I).

Nevertheless the slim-disk model,
as well as our model for the case of $\varepsilon_{100}=0.17$,
can reproduce
the basic trend of the ULXs data in the X-ray H-R diagram
(Watarai, Mizuno, \& Mineshige 2001).
However, we should stress that 
the physical reason to cause apparent decrease in $R_{\rm in}$
(or ratio of $L/T_{\rm eff}^4$) completely differ.
In the slim-disk approach,
the physical inner edge actually shifts and extends down to $\sim r_{\rm g}$
at high $\dot{m}$.
The consequence is that
the ULXs data can be fitted well
with small $R_{\rm in}$.
In our model, in contrast
the disk inner edge is assumed not change, but
the disk temperature increases apparently 
due to the transmission of the high-energy photons
from deep inside.
In realistic situations, both will work, but
which is more important is not clear at the moment.

Finally, we indicate the important difference between
the slim-disk model and our model.
The black-hole mass of IC 342 S1 is estimated to be $\sim 30M_\odot$
in the slim-disk approximation
(Watarai, Mizuno, \& Mineshige 2001),
whereas it is 
considerably massive, $50-100M_\odot$, in the present study.
We need to remark on the limitations of the present model,
however, since our model is essentially one-dimensional model
and the flow structure is not properly solved.
We need further studies (see below).

\subsection{Future Work}
Throughout the present study, 
we treated only the free-free absorption/emission and
the Thomson scattering as physical interaction
between gas and radiation.
However, the Comptonization may play an important role
on the emergent SED
(Czerny \& Elvis 1987; Ross, Fabian, \& Mineshige 1992;
Shimura \& Takahara 1993).
Then, the emergent spectra get even harder.
If the corona exists above the disk,
farther, the high-energy power-low photons will be generated 
by the inverse Compton scattering
(Haardt \& Maraschi 1991;
Liu, Mineshige, \& Ohsuga 2002).
The emergent SED would be improved by 
simulations of radiation transfer 
which incorporates Comptonization,
e.g., Monte Carlo simulations
(Pozdnyakov, Sobol', \& Sunyaev 1977).

The metal absorption may be effective 
at high frequency.
Then, the high-energy photons generated deep inside the disk
may be absorbed at a certain depth.
If this is the case, the higher-energy peak may disappear and
the peak of the emergent SED shift to lower-energy side.
We expect that the emission lines by metals should be 
observed in the emergent SED.
If the cross-section for the metal absorption is comparable to
or more than that of the Thomson scattering,
the photon trapping is drastically enhanced
and the energy-conversion efficiency is more reduced.
Then, such accretion flows will be observed as relatively dim
and metal rich objects.

In this study, we adopted the plane-parallel approximation,
which introduces error of $\sim (H/r)^2 \ll 1$.
As future work, we need to solve full two-dimensional radiation transfer
equations.

The convection
which seems to occur when the disk is radiation pressure dominated
(Shakura, Sunyaev, \& Zilitinkevich 1978;
Agol et al. 2001)
would also reduce the photon-trapping effects.
Since the high-energy photons
trapped in the flow 
can be carried towards the disk surface by the convection,
the high-energy photons emerge 
and the peak of the SED would be higher than
that of the expectation in our present study
(see discussion in Mineshige et al. 2000).

The photon-trapping effects
would be much reduced,
if the gas density distribution is strongly patchy in the disk,
since the photons can pass through the disk along the 
low-density regions.
Begelman (2002) suggested that by such a effect 
the disk luminosity 
can exceed the Eddington luminosity.
The details are not clear before multi-dimensional 
radiation-hydrodynamical (RHD) simulations are performed.
Two-dimensional RHD simulations
have been initiated by Eggum, Coroniti, \& Katz (1987, 1988),
and followed by 
Okuda, Fujita, \& Sakashita (1997), Fujita \& Okuda (1998),
Kley \& Lin (1999), Okuda \& Fujita (2000), and Okuda (2002),
though the radiation fields are treated in monochromatic form.

Finally, we employ the FLD approximation in the present work.
Though this formulation well agrees with the exact solution
both in the optically thick and thin regimes,
it would be better to solve the radiation transfer equations 
without using the approximation.
Especially, the FLD approximation might be invalid 
when the patchy structure forms inside the accretion flow,
since the FLD approximation does not always give 
the exact solution at the regions of moderate optical thickness.

\section{CONCLUSIONS}
By solving the multi-frequency radiation transfer and the energy equations of 
gas as well as radiation,
we have investigated the SED of the supercritical 
accretion flow 
with special attention on the photon-trapping effects.
The present results are summarized as follows.

(1)
For $L\lsim 2L_{\rm E}$ ($\dot{m}\lsim 30$), at which
the photon trapping does not occur,
the peak frequency of the resultant SED becomes,
at maximum, 
three times higher than that expected by the standard-disk theory.
This is a result of the transmission of the high-energy photons
from deep inside
because of smaller opacity for higher-energy radiation.
This result might resolve the so-called
\lq too hot accretion disk problem' (Makishima et al. 2000).

(2)
For $L\gsim 2L_{\rm E}$ ($\dot{m}\gsim 30$), 
the photon-trapping effects are substantial.
Then, 
the high-energy photons are more efficiently trapped in the flow
than the low-energy ones,
since the former arise near the equatorial plane.
Thus, the SED shifts towards the lower-energy side
as $\dot{m}$ increases.

(3)
By comparing with the observational data of ULXs
in the X-ray H-R diagram,
we found that the photon-trapping effects would only moderately 
appear in the accretion flow of ULXs.
We can explain the observed time variations of IC 342 S1
on the X-ray H-R diagram
in terms of $\dot{m}$ modulation. 
Its black-hole mass
is estimated to be $M\sim 100M_\odot$, which is greater than the previous
estimate based on the slim-disk model, $M\sim 30M_\odot$
(Watarai, Mizuno, \& Mineshige 2001).

\acknowledgments

The calculations were carried out at 
Yukawa Institute for Theoretical Physics in
Kyoto University.
This work is 
supported in part by Research Fellowships of the Japan Society
for the Promotion of Science for Young Scientists, 02796 (KO)
and 01680 (KW),
and the Grants-in Aid of the
Ministry of Education, Science, Culture, and Sport, 
13640238 (SM).


\begin{references}
\reference{}
Abramowicz, M. A., Czerny, B., Lasota, J. P., \& Szuszkiewicz, E. 
1988, ApJ, 332, 646
\reference{}
Agol, E., Krolik, J., Turner, N. J., \& Stone, J. M.
2001, ApJ, 558, 543
\reference{}
Alme, M. L. \& Wilson, J. R. 1976, ApJ, 210, 233
\reference{}
Begelman, M. C. 1978, MNRAS, 184, 53
\reference{}
Begelman, M. C. 2002, ApJ, 568, L97
\reference{}
Boller, T. 2000, NewAR, 44, 387
\reference{}
Colbert, E. J. M. \& Mushotzky, R. F. 1999, ApJ, 519, 89
\reference{}
Collin, S. et al.
2002, A\&A, 388, 771
\reference{}
Czerny, B. \& Elvis, M. 1987, ApJ, 321, 305
\reference{}
Eggum, G. E., Coroniti, F. V., \& Katz, J. I. 1987, ApJ, 323, 634
\reference{}
Eggum, G. E., Coroniti, F. V., \& Katz, J. I. 1988, ApJ, 330, 142
\reference{}
Frank, J., King, A., \& Raine, D. 1985,
Accretion Power in Astrophysics (Cambridge: Cambridge Univ. Press)
\reference{}
Fujita, M. \& Okuda, T. 1998, PASJ, 50 639
\reference{}
Fukue, J., Kato, S., \& Matsumoto, R. 1985, PASJ, 37, 383
\reference{}
Haardt, F. \& Maraschi, L. 1991, ApJ, 380, L51
\reference{}
Kato, S., Fukue, J., \& Mineshige, S. 1998,
Black-Hole Accretion Disks (Kyoto: Kyoto Univ. Press)
\reference{}
King, A. R., Frank, J., Kolb, U., \& Ritter, H. 1997, ApJ, 482, 919
\reference{}
Kley, W. \& Lin, D. N. C. 1999, ApJ, 518, 833
\reference{}
Kohri, K. \& Mineshige, S. 2002, ApJ, 577, 311
\reference{}
Laor, A. \& Netzer, H. 1989, MNRAS, 238, 897
\reference{}
Liu, B., Mineshige, S., \& Ohsuga, K. 2003, ApJ, in press
\reference{}
Lynden-Bell, D. \& Pringle, J. E. 1974, MNRAS, 168, 603
\reference{}
Makishima, K. et al.
      2000, ApJ, 535, 632
\reference{}
Mihalas, D. \& Klein, R. I. 1982, J. Comput. Phys., 46, 97
\reference{}
Mihalas, D. \& Mihalas, B. W. 1984, 
Foundations of Radiation Hydrodynamics (Oxford: Oxford Univ. Press)
\reference{}
Mineshige, S., Kawaguchi, T., Takeuchi, M., \& Hayashida, K. 2000,
PASJ, 52, 499
\reference{}
Mitsuda, K. et al. 1984, PASJ, 36, 741
\reference{}
Mizuno, T., Ohnishi, T., Kubota, A., Makishima, K., \& Tashiro, M. 
1999, PASJ, 51, 663
\reference{}
Narayan, R., Piran, T., \& Kumar, P. 2001, ApJ, 557, 949
\reference{}
Norman, C. \& Scoville, N. 1988, ApJ, 332, 124
\reference{}
Ohsuga, K., Mineshige, S., Mori, M., Umemura, M.
2002, ApJ, 574, 315 (Paper I) 
\reference{}
Ohsuga, K., Umemura, M., Fukue, J., \& Mineshige, S.
1999, PASJ, 51, 345
\reference{}
Okada, K., Dotani, T., Makishima, K., Mitsuda, K., \& Mihara, T. 
1998, PASJ, 50, 25
\reference{}
Okuda, T., Fujita, M., \& Sakashita, S. 1997, PASJ, 49, 679
\reference{}
Okuda, T. \& Fujita, M. 2000, PASJ, 52, L5 
\reference{}
Okuda, T. 2002, PASJ, 54, 253
\reference{}
Pozdnyakov, L. A.,  Sobol', I. M., \& Sunyaev, R. A.
1977, Soviet Astron., 21, 708
\reference{}
Revnivtsev, M., Gilfanov, M., Churazov, E., \& Sunyaev, R.
2002, A\&A, 391, 1013
\reference{}
Ross, R. R., Fabian, A. C., \& Mineshige, S. 1992, MNRAS,258,189
\reference{}
Rybicki, G. B. \& Lightman, A. P. 1979,
Radiative Processes in Astrophysics (New York: John Wiley \& Sons, Inc.)
\reference{}
Shakura, N. I. \& Sunyaev, R. A. 1973, A\&A, 24, 337
\reference{}
Shakura, N. I., Sunyaev, R. A., \& Zilitinkevich, S. S. 1978, A\&A, 62, 179
\reference{}
Shimura, T. \& Takahara, F. 1993, ApJ, 419, 78
\reference{}
Spruit, H. C. \& Ritter, H. 1983, A\&A, 124, 267
\reference{}
Stone, J. M., Mihalas, D., \& Norman, M. L. 1992, ApJS, 80, 819
\reference{}
Szuszkiewicz, E., Malkan, M. A., \& Abramowicz, M. A. 1996, ApJ, 458,
474
\reference{}
Turner, N. J. \& Stone, J. M. 2001, ApJS, 135, 95
\reference{}
Umemura, M., Fukue, J., \& Mineshige, S. 1998, MNRAS, 299, 1123
\reference{}
Wang, J. M., Szuszkiewicz, E., Lu, F. J., \& Zhou, Y. Y. 1999, ApJ,
522, 839
\reference{}
Watarai, K., Fukue, J., Takeuchi, M., \& Mineshige, S.
2000, PASJ, 52, 133
\reference{}
Watarai, K., Mizuno, T., \& Mineshige, S. 2001, ApJ, 549, L77
\reference{}
Yamaoka, K., Ueda, Y., Inoue, H. 2001, 
in ASP Conf. Ser. 251, New Century of X-Ray Astronomy, ed.
H. Inoue \& H. Kunieda (San Francisco: ASP)
\end{references}
\end{document}